\documentclass[times, 10pt,twocolumn]{article}
\usepackage{latex8}
\usepackage{graphicx}
\usepackage{mathrsfs}
\usepackage{amssymb}
\usepackage{amsmath}
\input{psfig.sty}

\newcommand{\prj}               {\mathit{\,prj\,\,}}
\newcommand{\CHOP}              {\boldsymbol{;} }
\newcommand{\IFF}               {\mathit{iff}}
\newcommand{\prob}              {\mathit{Prob}}
\newcommand{\true}              {\mathit{true}}

\newcommand{\interp}            {{\cal{I}}}
\newcommand{\emptyy}            {\varepsilon}

\newcommand{\DEF}               {\stackrel{\rm def}{=}}

\newtheorem{Expl}{Example}
\newtheorem{Def}{Definition}
\newtheorem{Thm}{Theorem}
\newtheorem{Lem}[Thm]{Lemma}

\def\squareforqed{\hbox{\rlap{$\sqcap$}$\sqcup$}}
\def\qed{\ifmmode\squareforqed\else{\unskip\nobreak\hfil
\penalty50\hskip1em\null\nobreak\hfil\squareforqed
\parfillskip=0pt\finalhyphendemerits=0\endgraf}\fi}

\begin{document}

       \title{A Probabilistic Variant of Projection Temporal Logic}
       \author{Xiaoxiao Yang   \\
           State Key Laboratory of Computer Science, \\
           Institute of Software, Chinese Academy of Sciences
            \\
            Beijing, 100190, China
            \\
            xxyang@ios.ac.cn
           }

       \maketitle

       \begin{abstract}
       In this paper, we propose Probabilistic discrete-time Projection Temporal Logic (PrPTL),
       which extends Projection Temporal Logic (PTL) with probability.
       To this end, some useful formulas are derived and some logic laws are given.
       Further, we define Time Normal Form (TNF) for PrPTL as the standard form  and prove that any PrPTL formulas
       can be rewritten to TNF. According to the TNF, we construct the time normal form graph which
       can be used for the probabilistic model checking on PrPTL.
       \end{abstract}

           \section{Introduction}
           In real-life systems, there are many phenomena
           that can be modeled by considering their stochastic characteristics. For this purpose,
           probabilistic model checking is proposed as a formal verification technique for the analysis of
           stochastic systems.  The probabilistic model checking problem is to compute the
           \emph{probability} for the set of paths in the model that satisfy a
           given property, which is based on quantitative logics and
           quantitative systems \cite{Katoen}. Properties to be analysed by probabilistic model checking can be
           formalized in some quantitative temporal logics such as
           probabilistic computation tree logic (PCTL) \cite{Hansson94} and continuous stochastic logic (CSL) \cite{Aziz00}.
           This paper investigates a new quantitative temporal
           logic, called Probabilistic discrete-time Projection Temporal Logic (PrPTL), which extends projection temporal
           logic (PTL) \cite{D96,DYK08,ZCL07} with probability and discrete time.

           Linear-time property is a set of infinite paths. We can use linear-time temporal logic (LTL)
           to express a subset of $\omega$-regular properties. However, PTL can specify more linear-time properties
           since the \emph{chop star} ($*$) and \emph{projection} operators in PTL are equivalent to the
           full $\omega$-regular languages. To investigate the probabilistic model checking based on PTL,
           we  propose PrPTL that can be used to specify quantitative linear-time properties.
           Further, we give the logic laws and derived formulas
           and prove that any PrPTL formulas can be reduced to a standard form called time normal form (TNF).
           In addition, according to the TNF, the model of PrPTL can be constructed, which is a basis for
           probabilistic model checking on PrPTL.


           \section{Projection Temporal Logic}

           Let $AP$ be a finite set of atomic propositions. Propositional PTL formulas over $AP$ can be defined as follows:

           \[
              Q ::= \pi \mid \neg Q \mid \bigcirc Q \mid  Q_1 \wedge Q_2  \mid (Q_1, \ldots, Q_m) \prj Q
           \]
           where $\pi \in AP$, $Q, Q_1,\ldots, Q_n$ are propositional PTL formulas, $\bigcirc$ (next) and $\prj$ (projection)  are basic temporal operators.
           A formula is called a \emph{state} formula if it does
           not contain any temporal operators, i.e.,  \emph{next} ($\bigcirc$),
           \emph{projection} ($\prj$); otherwise
           it is a \emph{temporal} formula.

           An interval $\sigma = \langle s_0,s_1,\ldots\rangle$ is a non-empty sequence of states,
           where $s_i~(i \geq 0)$ is a state mapping from $AP$ to $B= \{true, false\}$.
           The length, $|\sigma|$, of $\sigma$ is $\omega$ if $\sigma$ is infinite, and the number of states minus 1
           if $\sigma$ is finite.

           Let $N_0$ denote non-negative integers. An interpretation for a propositional PTL formula is a tuple $\interp =(\sigma,i,j)$, where $\sigma$ is an interval,
           $i$ is an integer, and $j$ is an integer or $\omega$ such that $i \leq j ~(i, j \in N_0$). Intuitively, $(\sigma,i,j)$
           means that a formula is interpreted over a subinterval
           $\sigma_{(i,..,j)}$. The satisfaction relation ($\models$) between interpretation $\interp$ and formula
           $Q$ is inductively defined as follows.
          \begin{center}
          \begin{enumerate}
                \item $\interp \models \pi$ iff $s_k[\pi]= true$
                \item $\interp \models \neg Q$ iff $\interp \nvDash Q$
                \item $\interp \models Q_1 \wedge Q_2$ iff $\interp \models
                Q_1$ and $\interp \models Q_2$
                \item $\interp \models \bigcirc Q$ iff $k < j$ and $(\sigma, i, k+1, j) \models Q$
                \item $\interp \models (Q_1,\ldots,Q_m) \prj Q$  iff there are~
                $k=r_0\leq r_1\leq\ldots \leq r_m \leq j$~\mbox{such that }
                $(\sigma,i,r_0,r_1)\models Q_1 ~\mbox{and}$
                $(\sigma,r_{l-1},r_{l-1}, r_l)\models Q_l~\mbox{for all}~1<l\leq m ~\mbox{and}$
                $ (\sigma',0,0,|\sigma'|)$ $\models Q ~\mbox{for}~ \sigma'~\mbox{given  by}:$
                \\
                $ (a) ~r_m<j ~\mbox{and}~ \sigma' = \sigma \downarrow (r_0,\ldots,r_m)~{\cdot}~\sigma_{(r_m+1,..,j)}$
                \\
                $(b) ~ r_m=j ~ \mbox{and}~ \sigma'=\sigma \downarrow (r_0,\ldots,r_h) ~\mbox{for some}~ 0\leq h\leq  m$.

        \end{enumerate}
        \end{center}

      \section{A Probabilistic Variant for PTL}
      Probabilistic discrete-time Projection Temporal Logic (PrPTL) is a quantitative variant
      of PTL.
      Based on the projection operator $(Q1, \ldots, Q_m) \prj Q$, we can define
      the sequential operator $P~\CHOP~Q$ as
      \[
         P~\CHOP~Q \DEF (P, Q) \prj \true
      \]
      which means that $P$
      holds from now until some point in future and from that time point
      $Q$ holds. For simplicity, we will employ sequential operator $P~\CHOP~Q$ instead of the projection operator
      to define PrPTL.

       \subsection{Syntax and Semantics}

      \begin{Def}
       The formulas in PrPTL are inductively defined as follows.
      \begin{eqnarray*}
       P &:: =& \pi \mid \neg P \mid P_1 \wedge P_2 \mid \bigcirc^{[t_1,t_2]} P
                       \mid  P_1 \CHOP^{[t_1,t_2]} P_2  \\
       \psi &::=& [P]_{\trianglelefteq p}
      \end{eqnarray*}
       where $\pi$ is an atomic proposition, $\bigcirc$ and $\CHOP$ are
       temporal operators, $p \in [0,1]$ is a probability, $\trianglelefteq \in \{<, \leq, \geq,>\}$,
       $t_1 \leq t_2 \in N_{\omega}~(N_{\omega} = N_0 \cup \omega )$ denotes time.
      \end{Def}

        \begin{enumerate}
       \item $ (\sigma,i,|\sigma|) \models \pi  ~~\IFF ~~\sigma(i) \models \pi $

       \item $(\sigma,i,|\sigma|) \models \neg P   ~~    \IFF ~~  (\sigma, i, |\sigma|) \not \models P $

       \item $(\sigma,i,|\sigma|) \models P_1 \wedge P_2   ~~   \IFF ~~  (\sigma, i, |\sigma|) \models P_1 \mbox{ and }  (\sigma, i, |\sigma|) \models P_2$

       \item $(\sigma,i,|\sigma|) \models \bigcirc^{[t_1,t_2]} P        ~~   \IFF  ~~ \exists ~l, t_1 \leq l \leq t_2, i+l \leq j,
       \mbox{ such that } (\sigma, i+l, |\sigma|) \models P $

       \item $(\sigma,i,|\sigma|) \models  P_1 \CHOP^{[t_1,t_2]} P_2      ~~ \IFF ~~   \exists ~r \leq |\sigma| \mbox{ such that }
       (\sigma, i, r) \models P_1 \mbox{ and } \exists ~l, t_1 \leq l \leq t_2, r+l \leq |\sigma|  \mbox{ such}~
       \mbox{that }(\sigma, r+l, |\sigma|) \models P_2 $

       \item $(\sigma, i, |\sigma|) \models \psi    ~~ \IFF ~~  \prob (\sigma_{(i..|\sigma|)}, P) \trianglelefteq p $
       \end{enumerate}

       As usual, $true \DEF P \vee \neg P$. If there is an
       interpretation $\interp$ such that $\interp\models P$ then a
       formula $P$ is \emph{satisfiable}. We also define the
       satisfaction relation for an interval $\sigma$ and formula $P$,
       by stating that $\sigma\models P$ if $(\sigma,0,|\sigma|)\models
       P$. Furthermore, we denote $\models P$ if $\sigma\models
       P$, for all intervals $\sigma$.


   For $t_1 = t_2 =t$, we abbreviate $[t,t]$ as $[t]$. Particularly, when $t_1=t_2=0$, $\bigcirc^{[0]} P$ denotes $P$ and $P_1
   \CHOP^{[0]} P_2$ denotes $P_1 ~\CHOP~ P_2$.
   Except the projection operator, all the basic formulas in propositional PTL can be defined in PrPTL.
   \[
   \begin{array}{lll}
   \bigcirc P     &\triangleq&    \bigcirc^{[1]} P \\
   \Diamond P     &\triangleq&    \true \CHOP P \\
   \Box P         &\triangleq&    \neg \Diamond \neg P     \\
   P_1 U P_2      &\triangleq&    P_1 \CHOP \bigcirc P_2 = P_1\CHOP^{[1]} P_2\\
   \emptyy        &\triangleq&    \neg \bigcirc \true\\
   more           &\triangleq&    \bigcirc true\\
   skip           &\triangleq&    \bigcirc \emptyy\\
   len(n)         &\triangleq&    \left\{
                                         \begin{array}{ll}
                                         \emptyy & \mbox{if} ~n=0\\
                                         \bigcirc len(n-1) & \mbox{if} ~n>1
                                         \end{array}
                                  \right.\\
   keep(P)        &\triangleq&    \Box (\neg \emptyy \rightarrow P)\\
   halt(P)        &\triangleq&    \Box (\emptyy \leftrightarrow P)\\
   fin(P)         &\triangleq&    \Box (\emptyy \rightarrow P)\\
   \Diamond^{[t_1,t_2]} P      &\triangleq&   \bigcirc^{[t_1,t_2]} P \\
   \Box^{[t_1,t_2]} P          &\triangleq&   \neg \Diamond^{[t_1,t_2]} \neg P     \\
   P_1 U^{\leq t} P_2          &\triangleq&   P_1 \vee (\Box^{<t}P_1\CHOP \bigcirc P_2)
   \end{array}
   \]

   \begin{Def}
   Two formulas, $P$ and $Q$, are equivalent, denoted $P \equiv Q$, if $\models \Box (P\leftrightarrow Q)$.
   \end{Def}

   Compared with the probabilistic computation tree logic (PCTL) \cite{Hansson94}, our logic can express more quantitative properties.
   Let $p$ and $q$ be atomic propositions. 
   Note that $p~U^{\leq 3}q$ in PCTL can be defined as $q \vee (\Box^{\leq 2} p ~\CHOP~\bigcirc
   q)$ in PrPTL.



     \subsection{Time Normal Form}
      We now give a standard form, called Time normal form, for PrPTL. 
      
     \begin{Def}
      Let $P$ be a PrPTL formula. Time normal form (TNF) of $P$ can be defined as
      \[
         P \equiv (\bigvee\limits_{i=1}^k  P_{e_i} \wedge \emptyy) \vee
                  (\bigvee\limits_{j=1}^h  P_{c_j}  \wedge \bigcirc^{[t_1,t_2]} P_{f_j} )
       \]
      where $k+h \geq 1, t_2 \geq t_1 \geq 1$, $P_{e_i}$ and $P_{c_j}$ are true or atomic propositions.
      \end{Def}

      For convenience,  we abbreviate
     $\bigvee\limits_{i=1}^k$ and $\bigvee\limits_{j=1}^h$ as $\bigvee$. Thus, TNF can be written as
     $P \equiv (\bigvee  P_{e} \wedge \emptyy) \vee (\bigvee P_{c}  \wedge \bigcirc^{[t_1,t_2]} P_{f} ) $.

     \begin{Lem}\label{laws-lem}
     Let $P$, $Q$ and $R$ be PrPTL formulas and $w$ a state formula. The following laws hold:
     \[
        \begin{array}{rlll}
        (L1)  & \bigcirc P \CHOP ^{[t_1,t_2]} Q &\equiv&  \bigcirc (P \CHOP^{[t_1,t_2]} Q)\\
        (L2)  & \emptyy \CHOP^{[t_1,t_2]} P     &\equiv&  \bigcirc^{[t_1,t_2]} P\\
        (L3)  & (w\wedge P) \CHOP^{[t_1,t_2]} Q &\equiv&  w \wedge (P \CHOP^{[t_1,t_2]}Q)\\
        (L4)  & \bigcirc^{[t_1,t_2]} P \wedge (Q \vee R) &\equiv& (\bigcirc^{[t_1,t_2]} P \wedge Q) \vee
        \\ &&& (\bigcirc^{[t_1,t_2]} P \wedge R)\\
        (L5)  & P \CHOP^{[t_1,t_2]} (Q \vee R)  &\equiv& (P \CHOP^{[t_1,t_2]} Q) \vee (P \CHOP^{[t_1,t_2]} R)
        \end{array}
     \]
     \end{Lem}

     \begin{Def}
     A Time Normal Form $P \equiv (\bigvee P_e \wedge \emptyy) \vee (\bigvee P_{j} \wedge \bigcirc^{[t_1,t_2]} P'_{j})$ for a PrPTL formula
     $P$ is called Complete Time Normal Form (CTNF) if
     \[ \bigvee \limits_j P_{j} \equiv true \mbox{ and } \bigvee \limits_{i \neq j} (P_{i} \wedge P_{j}) \equiv false\]
     \end{Def}

     \begin{Thm}
     For any formula $P$ in TNF, it can be rewritten into CTNF.
     \end{Thm}

     \begin{Thm}\label{TNF-thm}
     For any PrPTL formula $P$ there is a PrPTL formula $Q$ in TNF such that
                        \[   P \equiv Q  \]
     \end{Thm}


    \begin{Expl}

    Let $P$, $Q$ and $R$ be atomic propositions. The time normal form for formulas $P$
    and  $P ~\CHOP \bigcirc^{[3,4]} Q$ are reduced as follows.
     \begin{enumerate}
     \item TNF of $P$:
     \[
     P \equiv P \wedge true
       \equiv P \wedge (\bigvee \limits_{n=0}^{\omega} \bigcirc^n \emptyy)
       \equiv P \wedge \bigcirc^{[0, \omega]} \emptyy
     \]

     \item TNF of $P ~\CHOP \bigcirc^{[3,4]} Q$:
     \begin{eqnarray*}
     &&        P ~\CHOP \bigcirc^{[3,4]} Q \\
     &\equiv& ((P \wedge \emptyy) \vee (P \wedge \bigcirc true))\CHOP \bigcirc^{[3,4]} Q \\
     &\equiv& ((P \wedge \emptyy) \CHOP \bigcirc^{[3,4]} Q) \vee ((P \wedge \bigcirc true)\CHOP \bigcirc^{[3,4]} Q)\\
     &\equiv& (P  \wedge \bigcirc^{[3,4]} Q) \vee (P \wedge \bigcirc (true \CHOP \bigcirc^{[3,4]} Q))
     \end{eqnarray*}

    \end{enumerate}

\end{Expl}


 \subsection{Time Normal Form Graph}

 It is proved that any PrPTL formula $P$ can be rewritten into TNF.
 Based on TNF, we now construct a model called Time Normal Form Graph (TNFG) for PrPTL.
 Tuple $(Q,T)$ denotes that holding formula $Q$ for time $T$. When $T=0$, we often omit the time, and write $(Q, 0)$ as $Q$.

 \begin{Def}\label{def-TNFG1}\rm
 For PrPTL formula $P$, let $V(P)$ be a set of nodes and $E(P)$ be a set of edges.
 Graph $G=(V(P),E(P))$ is defined as follows.

 \begin{itemize}
 \item  $P\in V(P)$;

 \item For all $(Q,T) \in V(P)$, if $Q \equiv (\bigvee \limits_{i=1}^k Q_{ei} \wedge \emptyy) \vee (\bigvee\limits_{j=1}^h Q_{cj} \wedge \bigcirc^{[t_1,t_2]} Q_j)$,
       then $\emptyy \in V(P)$, $((Q,T), Q_{ei}, \emptyy ) \in E(P)$ for each $i, 1 \leq i \leq k$; $(Q_j, [t_1-1,t_2-1]) \in V(P)$,
       $((Q,T), Q_{cj}, (Q_j,[t_1-1,t_2-1]) \in E(P)$ for each $j, 1 \leq j \leq h$.
 \end{itemize}
 \end{Def}

 \begin{Def}\label{def-TNFG2}\rm

  TNFG of $P$ is a directed graph $G' = (G, \textbf{P})$, where \textbf{P}: $E(P) \rightarrow [0,1]$ is a probability.
 \end{Def}
 
 \begin{Expl}
  Let $P$ and $Q$ be atomic propositions. TNFG of formulas $P$, $P~\CHOP~ \bigcirc^{[4,5]}Q$ and $[P]_{=0.5}$ are shown as follows.

\begin{figure}[htpb]
  \includegraphics[width=8.2cm]{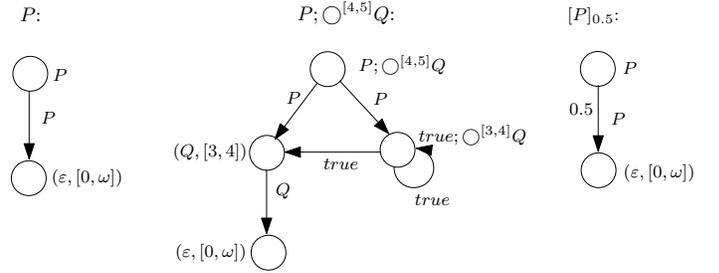}
  \caption{Examples of TNFG.}\label{model-eg1}
\end{figure}
 \end{Expl}


\section{Conclusion}

           This paper presents a probabilistic variant of projection temporal logic, PrPTL.
           The time normal form is defined and some logic laws are given. Then TNFG for capturing the models of PrPTL formulas is constructed. 
           In the near future, we will extend the existing model checker for propositional PTL with probability,
           and according to the TNFG proposed in this paper
           to verify the quantitative linear-time properties in probabilistic systems.

\bibliographystyle{latex8}

\bibliography{latex8}

\end{document}